\documentstyle[11pt,aaspp4]{article}

\newcommand{\AAA}{{\rm \AA}}
\newcommand{\OII}{[O\,II]}

\newcommand{\apriori}{{\em a priori\/}}

\newcommand{\ie}{{\em i.e.}}
\newcommand{\Ie}{{\em I.e.}}

\newcommand{\etal}{{\em et al.}}
\newcommand{\etc}{{\em etc.}}

\begin{document}

\title{
	The oxygen-II luminosity density of the Universe\altaffilmark{1}
}
\author{
	David W. Hogg\altaffilmark{2,3,4,5},
	Judith G. Cohen\altaffilmark{3},
	Roger Blandford\altaffilmark{2},
	Michael A. Pahre\altaffilmark{3,5,6}
}
\altaffiltext{1}{
Based on observations made at the W. M. Keck Observatory, which is
operated jointly by the California Institute of Technology and the
University of California; at the Palomar Observatory, which is owned
and operated by the California Institute of Technology; and with the
NASA/ESA Hubble Space Telescope, which is operated by AURA under NASA
contract NAS 5-26555.
}
\altaffiltext{2}{
Theoretical Astrophysics, California Institute of Technology,
mail code 130-33, Pasadena CA 91125, USA;
{\tt rdb@tapir.caltech.edu}
}
\altaffiltext{3}{
Palomar Observatory, California Institute of Technology,
mail code 105-24, Pasadena CA 91125, USA;
{\tt jlc@astro.caltech.edu}
}
\altaffiltext{4}{
School of Natural Sciences, Institute for Advanced Study,
Olden Lane, Princeton NJ 08540, USA;
{\tt hogg@ias.edu}
}
\altaffiltext{5}{
Hubble Fellow
}
\altaffiltext{6}{
Present address:\ 
Harvard-Smithsonian Center for Astrophysics, MS 20,
60 Garden St, Cambridge MA 02138, USA;
{\tt mpahre@cfa.harvard.edu}
}

\begin{abstract}
Equivalent widths of \OII\ 3727\,\AA\ lines are measured in 375 faint
galaxy spectra taken as part of the Caltech Faint Galaxy Redshift
Survey centered on the Hubble Deep Field.  The sensitivity of the
survey spectra to the \OII\ line is computed as a function of
magnitude, color and redshift.  The luminosity function of galaxies in
the \OII\ line and the integrated luminosity density of the Universe
in the \OII\ line are computed as a function of redshift.  It is found
that the luminosity density in the \OII\ line was a factor of $\sim
10$ higher at redshifts $z\sim 1$ than it is at the present day.  The
simplest interpretation is that the star formation rate density of the
Universe has declined dramatically since $z\sim 1$.
\end{abstract}


\section{Introduction}
\label{sec:cintro}

The \OII\ line at 3727\,\AA\ is actually a pair of atomic transitions
for singly ionized oxygen, the $^2D_{3/2}$ to $^4S_{3/2}$ at
3726\,\AA\ and the $^2D_{5/2}$ to $^4S_{3/2}$ at 3729\,\AA.  The
transitions are forbidden, meaning that there is no electric dipole
connection between the initial and final states, so the spontaneous
emission rates are small, $1.8\times10^{-4}$ and
$3.6\times10^{-5}~{\rm s^{-1}}$ for the 3726 and 3729\,\AA\
transitions respectively (Osterbrock 1989).  For this reason, the
\OII\ line is usually collisionally excited by free electrons in hot
nebulae; temperatures $T\sim 10^4$~K are needed to excite the 3.3~eV
transitions.  If the electron density is very low, collisional
excitation is rare, whereas if it is very high, excited atoms are more
likely to be deexcited by a subsequent collision than by spontaneous
emission, so there are critical electron densities $n_{\rm c}$ at
which the transitions saturate observationally, defined to be the
electron densities at which the collisional excitation rates equal the
spontaneous emission rates.  The critical densities depend on
temperature because the collisional excitation cross sections do, but
at typical temperatures they are roughly $1.6\times10^4$ and
$3\times10^3~{\rm cm^{-3}}$ for the 3726 and 3729\,\AA\ transitions
respectively (Osterbrock 1989).  (The fact that the two critical
densities are different means that the 3726/3729 line ratio can be
used to measure electron density.)

The conditions of temperature and density required to excite the \OII\
3727\,\AA\ line are met in H\,II regions, clouds of ionized hydrogen
heated by massive, young, luminous stars.  For this reason, the \OII\
emission of a galaxy is sensitive to its young stellar population, or
recent star formation history.  In the local Universe, the
relationship between \OII\ luminosity and star formation has been
calibrated
\begin{equation}
\frac{L_{\rm\OII}}{2\times10^{33}~{\rm W}} =
  \frac{R}{1~{\rm M_{\odot}\,yr^{-1}}}
\label{eq:calib}
\end{equation}
where $R$ is the star formation rate (Kennicutt 1992).  This
relationship shows a significant galaxy-to-galaxy scatter.  It depends
on galaxy dust content because dust absorbs strongly in the
ultraviolet; the stellar initial mass function because the \OII\
luminosity is tied only to the massive star population; and
metallicity because the luminosity in the optically thin line ought to
be proportional to oxygen abundance, which in turn depends on a
galaxy's age and star formation history.

In this study the \OII\ luminosity function and \OII\ luminosity
density of the Universe are measured as a function of redshift.  These
functions constrain the star formation history of the Universe.
Previous studies on the star formation history of the Universe have
used the metal abundances in quasar absorption systems (Pei \& Fall
1995) and broadband luminosity density (Lilly \etal\ 1996); both of
these studies show a much higher star formation rate at $z\sim 1$ than
at the present epoch.  The star formation rate of the local Universe
has been estimated through the luminosity density in the H$\alpha$
line (Gallego et al 1995), which is a good measure of star formation
rate (Kennicutt 1992).  However, at redshifts $z>0.3$ the H$\alpha$
line is no longer accessible by visual spectroscopy and therefore
difficult to measure.  The \OII\ line is a less reliable measure of
star formation rate (Kennicutt 1992) but it has the great advantage
that it can be measured by visual spectroscopy over the interesting
redshift range $0.3<z<1.3$ where the star formation rate density is
thought to be changing rapidly.  These considerations motivate the
present work.  Several recent studies have shown that the incidence of
strong \OII\ emission among bright galaxies increases with redshift
(Cowie et al 1996; Heyl et al 1997; Small et al 1997; Hammer et al
1997).  This trend implies a higher \OII\ luminosity density in the
past, which has also been previously measured directly (Hammer et al
1997).

In what follows, physical quantities are quoted in SI units, with
Hubble constant $H_0=100\,h~{\rm km\,s^{-1}\,Mpc^{-1}}$, in world
model $(\Omega_M,\Omega_{\Lambda})=(0.3,0)$.  The only exception are
number densities, which are given in $h^3\,{\rm Mpc^{-3}}$.  Fluxes
and luminosities are given as flux and luminosity densities per
logarithmic frequency interval, \ie, $\nu\,S_{\nu}$ or
$\lambda\,S_{\lambda}$ and $\nu\,L_{\nu}$ or $\lambda\,L_{\lambda}$,
in ${\rm W\,m^{-2}}$ and ${\rm W}$.  Luminosities are all-sphere (not
per-steradian).

\section{Sample, observations, and line measurement}
\label{sec:ewsample}

The galaxy sample utilized for this study is an incompletely observed
magnitude-limited sample, selected in the $\cal R$ band, in the Hubble
Deep Field (HDF, Williams \etal\ 1996) and an 8-arcmin diameter
circular field surrounding it.  The general sample selection,
photometry and main spectroscopic results are described elsewhere
(Hogg \etal\ in preparation; Cohen \etal\ in preparation).  Briefly,
the sample is selected to be all sources, independent of morphology,
brighter than ${\cal R}=23.3$~mag in the 8-arcmin diameter circular
field and brighter than ${\cal R}=24.5$~mag in the small HDF proper
(most of the central $2\times 2~{\rm arcmin}^2$).  The spectroscopy of
this sample is only about 75~percent complete for the purposes of this
study, which is based on 375 spectra.  (The redshift survey is more
than 90~percent complete but some of the redshifts come from spectra
taken by other groups.)  Fluxes in the $\cal R$ band and ${\cal
R}-K_s$ colors are measured with data from the COSMIC and Prime Focus
IR cameras on the Hale 200-inch Telescope (Kells et al 1998).
Spectroscopy is performed with the LRIS instrument on the Keck
Telescope (Oke \etal\ 1995) with a $300~{\rm mm}^{-1}$ grating, at a
resolution of about 10\,\AA\ (2.5\,\AA\ per pixel, 1~arcsec slits),
for exposure times of 6000 to 9000~s (Cohen \etal\ in preparation).
Figure~\ref{fig:eqwpretty} shows some example spectra from the sample,
cut out around the \OII\ 3727\,\AA\ line.

The continua are fit with a straight line over the wavelength range
from 200\,\AA\ to 50\,\AA\ shortward of the observed 3727 location and
the range from 50\,\AA\ to 200\,\AA\ longward.  Each fit is performed
with six iterations of sigma-clipping at $\pm 2.5\,\sigma$, where
$\sigma$ is the root-mean-square (RMS) residual noise per pixel.  The
uncertainty in the continuum value at the line center is taken to be
the per-pixel RMS divided by the square root of the number of pixels
contributing to the continuum fit (after sigma-clipping).

The line strength is measured by summing the differences between
observed spectrum and continuum fit in the 30\,\AA\ (full-width)
aperture centered on the line location.  The uncertainty in this
strength is taken to be the per-pixel RMS times the square root of the
number of pixels contributing to the line flux.

\section{Equivalent width distributions}

The rest-frame equivalent width $W$ of a line in the spectrum of an
object at redshift $z$ is the wavelength interval of continuum which
would provide the same total flux, corrected for redshift
\begin{equation}
W\equiv \frac{1}{1+z}\,\frac{\int \left[S_{\lambda}-S^{\rm(c)}_{\lambda}\right]\,d\lambda}{S^{\rm(c)}_{\lambda}}
\end{equation}
where $z$ is redshift, the integral is over only that spectral region
which contains the line, $S_{\lambda}$ is the flux density (per unit
observed wavelength $\lambda$) and $S^{\rm(c)}_{\lambda}$ is the flux
density in the continuum at the location of the line; \ie, the flux
density which would be observed in the absence of the line.  The
equivalent width is a robust measure of the strength of a spectral
feature relative to the source's continuum measure; it does not depend
on absolute calibration of the spectrum or even the relative
calibration of different parts of the spectrum, as long as spectral
response varies sufficiently slowly.  It is a local, geometric measure
of the line strength.

For the purposes of this study, the fractional uncertainty in an
equivalent width measurement is taken to be the sum in quadrature of
the fractional uncertainties in the continuum measurement and line
strength, both described in the previous section.  At low continuum
signal-to-noise ratios, this is not strictly correct because the
equivalent width error distribution is not gaussian or even symmetric
around the measured value.

The rest-frame \OII\ 3727 equivalent widths for the sample are shown
in Figures~\ref{fig:ewz} and \ref{fig:ewR}, plotted against $R$-band
magnitude and redshift $z$.  Only spectra with good (signal-to-noise
better than 2 in a pixel) continuum detections are plotted because
badly estimated or zero continuum leads to large, unreliable
equivalent width estimates.  Figures~\ref{fig:ewz} and \ref{fig:ewR}
are encouraging for those undertaking faint galaxy redshift surveys
because it shows that at higher redshifts and fainter fluxes, the
equivalent widths of \OII\ 3727 lines become greater.  (The observed
equivalent widths become even greater because of the additional factor
of $[1+z]$.)  Figures~\ref{fig:ewz} and \ref{fig:ewR} are subject to
an important selection effect: faint or high-redshift sources with
small equivalent widths may simply not be successfully assigned a
redshift at all.  This selection effect can clear out the
faint--small-width and high-redshift--small-width parts of these
diagrams.  However, the incompleteness in redshift identification is
less than 10~percent for this sample (Cohen \etal\ in preparation).
Furthermore, this selection effect does not explain any lack of
observed objects at bright levels or low redshifts with large
equivalent widths.  One recent study shows a strong inverse
correlation between galaxy luminosity and \OII\ equivalent width
(Cowie \etal\ 1996).  That inverse correlation is not seen as strongly
in the sample analyzed here, nor that of Small \etal\ (1997),
especially when sources with low signal-to-noise continuum
measurements are excluded (Small, private communication), as they have
been in Figures~\ref{fig:ewz} and \ref{fig:ewR}.  Because the
equivalent width is a ratio of observed quantities, there is a
tendency to overestimate the equivalent widths of lines in sources
with low signal-to-noise continuum measurements.

Figure~\ref{fig:ewz} may show evidence for clumps in
redshift--equivalent-width space.  The survey field is only 8~arcmin
in diameter, just a few Mpc at high redshift in typical models, so
sources with similar redshifts are likely to be physically associated.
This suggests that the galaxies which reside in the same high-redshift
group may also be related in terms of stellar content.  It suggests
that at least some of the galaxies in each group formed at the same
time and with similar stellar populations.  This is nicely consistent
with the observation that groups are long-lived, primordial structures
which exist at high redshift in relatively high abundance (Cohen
\etal\ 1996a, 1996b; Steidel \etal\ 1997).

\section{Sensitivity to line emission}
\label{sec:instsens}

The identification of the line at redshift $z$ depends on (a)~the
fraction of spectra in the sample which include wavelength
$(3727\,\AAA)(1+z)$ in their spectral range, (b)~the total sensitivity
of the atmosphere plus telescope plus instrument to line flux at
$(3727\,\AAA)(1+z)$, and (c)~the accuracy to which night sky and other
background emission can be subtracted at $(3727\,\AAA)(1+z)$.  Because
the spectrograph is a multi-slit design, different sources in the
survey are observed over different wavelength ranges, depending on the
position of the source within the field of the instrument.  The
wavelength coverage function can be constructed by taking the minimum
and maximum possible source locations and assuming that on any given
slitmask sources are evenly distributed between these extremes.  The
sensitivity to flux (in the sense of $\nu S_{\nu}$) can be estimated
with observations of spectrophotometric standard stars.  The
sensitivity varies from night to night, so in principle this function
should be replaced with a distribution function which takes into
account the variation in observing conditions.  Furthermore, in the
multislit design, if there are any positional errors in the catalog or
mask misalignment while observing, different sources will be centered
on their slits with different precisions.  This leads to a random
scatter in throughputs, even for sources observed simultaneously.  The
sky brightness, color and emission line spectrum also vary from night
to night.  In principle the expected sensitivity to 3727 emission can
be estimated from the coverage, sensitivity, and sky brightness
functions.  However, because the sensitivity depends on data reduction
technique, includes the complications of assessing slitmask alignment,
and may be compromised by unknown instrumental effects, a purely
empirical approach is taken here, using the reduced spectra themselves
to assess the sensitivity.

The signal-to-noise ratio $r$ (defined to be continuum level divided
by pixel-to-pixel rms), is measured in every spectrum in the sample at
a set of wavelengths corresponding to the 3727 line at various
redshifts in the range $0<z<1.8$, by exactly the procedure used to
estimate the continuum in the equivalent width measurements described
above.  The rms is computed from only those pixels not rejected by the
sigma-clipping algorithm, which is perhaps optimistic.  These
continuum signal-to-noise ratios are ``scaled'' to the value they
would have if the source had ${\cal R}=23$~mag and a spectral exponent
$n=0$ (the spectral exponent $n$ used in this work is defined
by $\nu\,S_{\nu}\propto\nu^n$ and it is measured from the ${\cal
R}-K_s$ color).  This scaling is done by multiplying the measured
signal-to-noise ratio by
\begin{eqnarray}
10^{0.4\,({\cal R}-23)}\,\left(\frac{1+z}{1.85}\right)^n
 & {\rm if} & {\cal R}>21.5~{\rm mag} \nonumber \\
10^{0.4\,(-1.5)}\,\left(\frac{1+z}{1.85}\right)^n
 & {\rm if} & {\cal R}<21.5~{\rm mag}
\label{eq:contfactor}
\end{eqnarray}
where the switch-over at ${\cal R}=21.5$~mag takes place because
at fainter magnitudes most of the source is in the slit and intensity
through the slit is proportional to total source flux, while brighter
than that the source is typically larger than the slit and intensity
through the slit depends only weakly on total source flux since the
bright galaxies in the sample have similar surface brightnesses.  The
$1+z$ term is divided by 1.85 because $z=0.85$ puts 3727 into the
center of the $\cal R$ band.  The switch-over magnitude was determined
by trial-and-error, with the test being that the distribution of
scaled signal-to-noise ratios not depend strongly on magnitude.  The
signal-to-noise ratio can be converted into a sensitivity to
rest-frame equivalent width, expressed in terms of the smallest
detectable rest-frame equivalent width
\begin{equation}
W_{\rm lim} = \frac{\eta\,\lambda_1}{r}
 \,\left(\frac{\Delta\lambda}{\lambda_1}\right)^{1/2}\,(1+z)^{-1/2}
\end{equation}
where $\eta$ is the minimum necessary signal-to-noise ratio for 3727
to be detected, taken to be 3, $r$ is the scaled signal-to-noise ratio
in the continuum, $\lambda_1$ is the wavelength per pixel, usually
2.5\,\AA\ for these spectra, $\Delta\lambda$ is the rest-frame
full-width of the 3727 line, taken to be 10\,\AA, and $z$ is the
redshift.  Because the formula for $W_{\rm lim}$ includes $r$ in the
denominator, the scaled value can be converted back into the true
sensitivity to rest-frame equivalent width by multiplying by the
factors given in (\ref{eq:contfactor}).

Since the continuum of every spectrum is measured at every redshift,
there are a huge number of scaled $W_{\rm lim}$ estimates from which a
model of the spectrograph sensitivity can be constructed.  At each
redshift the scaled sensitivities are ranked and a cumulative
distribution is constructed.  This distribution is shown in
Figure~\ref{fig:cumulativez}.  The distribution is plotted
cumulatively so that it can be treated as a probability, given a
source with a given redshift and \OII\ equivalent width, that the line
is detected.

Note that this sensitivity function is empirical, derived from the
sample of spectra themselves, and is only valid for this survey,
because it depends on the instrument, site, observational technique,
reduction method, and selection function.  The sensitivity becomes
worse at redshifts $1<z<1.25$ because the CCD efficiency at the
relevant wavelength is dropping while the sky brightness and number of
bright night sky emission lines are both increasing, and then very bad
at redshifts $1.25<z<1.5$ because in addition the fraction of spectra
with coverage at long enough wavelength is also decreasing.
Similarly, the bad sensitivity to \OII\ emission at low redshifts
$z<0.3$ is also caused by a decreasing fraction of spectra with
coverage at short enough wavelengths (although at these low redshifts
other spectral features can be used to determine redshifts).

\section{The \OII\ luminosity function}
\label{sec:o2glf}

For any galaxy, the line luminosity $L_{\rm\OII}$ can be crudely
computed with the rest-frame equivalent width $W$, the flux $S$
(defined to be $\nu\,S_{\nu}$) and the spectral exponent $n$ (defined
so $\nu\,S_{\nu}\propto\nu^n$) by
\begin{equation}
\log L_{\rm\OII} = \log\left[\frac{W}{3727\,\AAA}\right] + \log S
 + \log [4\,\pi] + 2\,\log D_L(z)
 - n\,\log \left[(1+z)\,\frac{3727\,\AAA}{\lambda_{\cal R}}\right]
\end{equation}
where $D_L(z)$ is the luminosity distance in an
$(\Omega_M,\Omega_{\Lambda})=(0.3,0.0)$ universe, and $\lambda_{\cal
R}$ is the effective wavelength of the $\cal R$ band, or 6900\,\AA.
Note that this is an all-sphere (not per-steradian) luminosity
definition.  Fluxes are derived from $\cal R$-band magnitudes using a
standard absolute calibration (Steidel \& Hamilton 1993).  This
prescription for line luminosity is crude because the spectral energy
distributions of galaxies are not pure power-laws, and, furthermore,
in this study, the spectral exponent $n$ has been computed from the
${\cal R}-K_s$ color, which does not ``bracket'' the \OII\ line unless
the redshift is $z>0.85$.  A refinement would be to compute $n$ from,
say, $G-{\cal R}$ at redshifts $z<0.85$.  In principle the need to use
the flux $S$ and exponent $n$ can be obviated entirely because line
fluxes can be measured directly from spectrophotometric data.
However, such procedures depend on perfect slit alignment on the
galaxies and aperture corrections to account for line flux outside the
slit.  The procedure used here is more robust.

The luminosity function is estimated with a modified version of the
$<V/V_{\rm max}>$ method, in which each galaxy in the survey is
assigned a volume $V_{\rm max}$ which is the volume of the Universe in
which that source could lie and still meet the survey criteria.  The
inverse volumes of all the galaxies in a particular luminosity bin are
summed to estimate the luminosity function in that bin.  In this
application, there are two important complications in computing
$V_{\rm max}$.  The first is that the survey is incomplete, in the
sense that only about 75~percent of the sources in the field are
observed as part of the subsample used here.
Figure~\ref{fig:o2pricomp} shows the \apriori\ completeness function,
which is defined to be the fraction of the total sources in the field
which were observed spectroscopically, as a function of $\cal R$-band
flux.  The second complication is that whether or not a source is in
the sample depends not only on the \apriori\ completeness function but
also on the detection of the \OII\ line itself, both because if it is
not detected there is no luminosity (for the luminosity function) and
because redshift identification often depends on \OII\ detection
anyway.  Fortunately, however, the sensitivity to the \OII\ line is
computed in Section~\ref{sec:instsens} and shown in
Figure~\ref{fig:cumulativez}.  Recall that the plotted sensitivity
function is scaled to an equivalent ${\cal R}=23$~mag, $n=0$ source by
the scaling given in (\ref{eq:contfactor}); the scaling and the
function in Figure~\ref{fig:cumulativez} can be combined to make a
total probability $p_{\rm detect}(S,n,z,W)$ of detecting an \OII\ line
of equivalent width $W$ in a source with flux $S$, spectral exponent
$n$ and redshift $z$.

Given the completeness function and detection probability function,
the appropriate formula for each galaxy's volume $V_{\rm max}$ is
\begin{equation}
V_{\rm max}=\int_0^{\infty}\eta_{\rm try}(S')
  \,p_{\rm detect}(S',n,z',W)\,\frac{d^2V_{z'}}{d\Omega\,dz'}
  \,\Delta\Omega\,dz'
\end{equation}
where $\eta_{\rm try}(S)$ is the probability that a spectrum was taken
of a source of flux $S$ in an attempt to get its redshift, and $S'$ is
the flux the source would have if it were at redshift $z'$ rather than
its true redshift.  The function $\eta_{\rm try}$ is plotted for this
spectroscopic sample in Figure~\ref{fig:o2pricomp}.  The luminosity
function $\phi(\log L_i)$ (number density per logarithmic interval in
luminosity) in a bin of \OII\ luminosity width $\Delta\log L$ centered
on \OII\ luminosity $L_{\rm\OII}=L_i$ is estimated with
\begin{equation}
\phi(\log L_i) = \frac{1}{\Delta(\log L)}
 \,\sum_{|\log L_{{\rm\OII},j}- \log L_i|<\Delta(\log L)/2}
   \frac{1}{V_{{\rm max},j}}
\end{equation}
where the sum is over all galaxies with luminosities in the bin, so
index $i$ labels luminosity bins and index $j$ labels galaxies.
Variances are computed by summing the squares of the inverse volumes;
the error bars on the Figures are the root variances.

The \OII\ luminosity function is shown in Figure~\ref{fig:o2vmax0} for
the entire sample, in the redshift range $0<z<1.5$.  It is compared to
the local H$\alpha$ luminosity function from the UCM survey (Gallego
\etal\ 1995) where the H$\alpha$ points have been multiplied by a
factor of 0.46, the mean observed \OII/H$\alpha$ flux ratio in the
local Universe (Kennicutt 1992).  Figure~\ref{fig:o2vmaxz} shows the
luminosity for two subsamples split in redshift at $z=0.35$.  This
Figure shows a strong evolution in the \OII\ luminosity function at
the bright end.  Although the total number density of \OII -emitting
galaxies is not significantly different between the two subsamples,
the typical line luminosity is higher by an order of magnitude in the
higher-redshift subsample.  Both subsamples show a higher line
luminosity than that which would be predicted from the very local UCM
results, given the local \OII/H$\alpha$ flux ratio.  Although there
may be some bias against luminous, low-redshift sources due to
undersampling, it is not strong enough to produce the apparent
evolution shown in Figure~\ref{fig:o2vmaxz}, especially since (a)~the
undersampling is accounted-for with the \apriori\ completeness
function and (b)~even the ``low-redshift'' sample goes to redshift
$z=0.35$, where there are many galaxies in the sample with luminosities
around $L^{\ast}$.

\section{The \OII\ luminosity and star formation rate densities}

As discussed in Section~\ref{sec:cintro}, the \OII\ line luminosity is
a star formation indicator, so the \OII\ luminosity function is a
measure of the star formation rate density of the Universe.  For these
purposes the entire luminosity function is not necessary, only the
integrated luminosity density is needed.  Because this is a single
number rather than a function, it is possible to subdivide the sample
more finely in redshift than was possible in Section~\ref{sec:o2glf}.

The luminosity density ${\cal L}_{\rm\OII}$ in the \OII\ line is
estimated similarly to the luminosity function, using the same volumes
$V_{\rm max}$ computed for those purposes.  The integrated luminosity
density is computed with
\begin{equation}
{\cal L}_{\rm\OII}= \,\sum_j \frac{L_{{\rm\OII},j}}{V_{{\rm max},j}}
\end{equation}
where galaxies are labeled by index $j$.  The variance on this
quantity is taken to be the Poisson value: the sum of the square
contributions.  The \OII\ line luminosity density as a function of
redshift, computed in two overlapping ``binnings,'' is shown in
Figure~\ref{fig:o2density}, along with the local measurement of the
H$\alpha$ luminosity density from the UCM survey, again scaled by the
local {\OII}/H$\alpha$ flux ratio.  The random errors on the
individual points in Figure~\ref{fig:o2density} are in fact expected
to be larger than the plotted poissonian uncertainties on account of
the strong redshift clustering found in this and other small redshift
survey fields (Cohen et al 1996b), in which more than 50~percent of
sources are found to be concentrated into a few narrow redshift spikes
out to $z\approx 1$.  This is seen dramatically in the point at
$z=0.5$, which is high even relative to its overlapping neighbors;
there are several large redshift overdensities in this $0.4<z<0.6$
bin.  Unfortunately this is a single-field study and only when several
independent fields have been similarly analyzed will it be possible to
average out such field-to-field variations.

After accounting for differences in Hubble constant and world model,
the luminosity density measurements shown in
Figure~\ref{fig:o2density} are in good agreement with those of Hammer
et al (1997).

Figure~\ref{fig:o2density} also shows the star formation rate density,
computed from the luminosity density with the Kennicutt (1992) local
calibration given by (\ref{eq:calib}).  Overall,
Figure~\ref{fig:o2density} implies that the star formation rate
density was nearly ten times higher at $z\sim 1$ than at the present
day.  A full analysis must take account of the changing metal, gas,
and dust contents of high-redshift galaxies, factors that are
difficult to assess with confidence at the present time.  Our results
are consistent with star formation rate density estimates based on
broadband luminosity density (Lilly et al 1996) and quasar absorption
line metallicities (Pei \& Fall 1995), both of which suggest
factor-of-ten reductions from $z\sim 1$ to the present day.

\acknowledgements The authors are pleased to thank Kurt Adelberger and
Chuck Stiedel, who provided some of the imaging data used in this
study, and Todd Small, Ed Groth, and the referee, David Spergel, for
helpful comments on the manuscript.  This research is based on
observations made at the W. M. Keck Observatory, which is operated
jointly by the California Institute of Technology and the University
of California; at the Palomar Observatory, which is owned and operated
by the California Institute of Technology; and with the NASA/ESA
Hubble Space Telescope, which is operated by AURA under NASA contract
NAS~5-26555.  Financial support was provided under NSF grant
AST~95-29170, Hubble Space Telescope archival research grant
AR-06337.12-94A, and Hubble Fellowship grants HF-01093.01-97A and
HF-01099.01-97A.  The Hubble grants are provided by STScI, which is
operated by AURA under NASA contract NAS~5-26555.  This research made
use of the NASA ADS Abstract Service and the SM software package.


\clearpage
\plotone{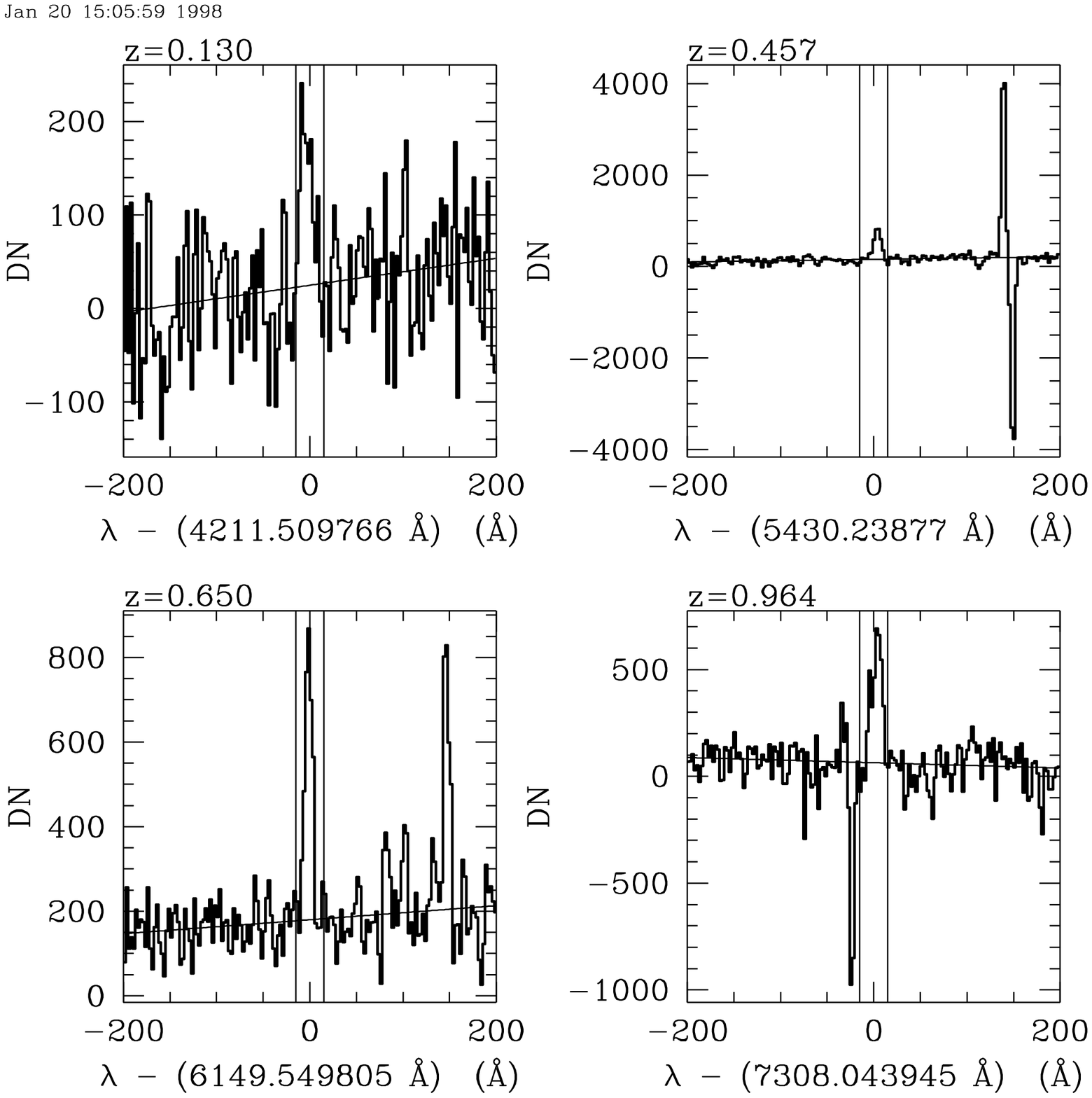}
\figcaption{
Example \OII\ 3727\,\AA\ line detections for four sources from the
sample.  The data (in ``data numbers'' or DN) are shown with a dark
line, the fit continuum with a thin straight line, and the aperture in
which the 3727 line strength is measured with two thin vertical lines.
The redshift of each source is given in the top left corner of each
plot.  Spikes or features not at zero wavelength are residuals of sky
lines imperfectly subtracted.
\label{fig:eqwpretty}}

\plotone{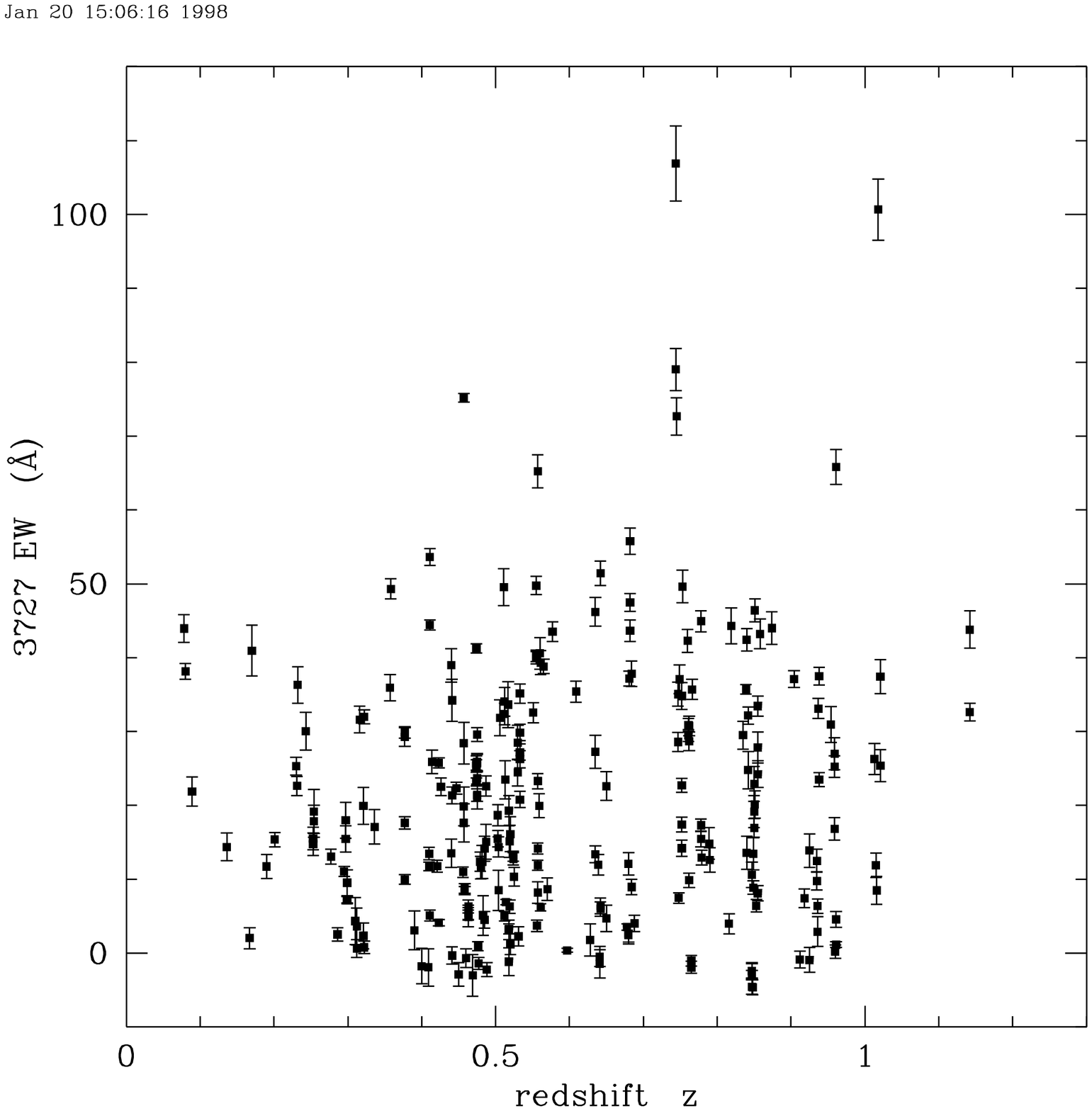}
\figcaption{
Rest-frame \OII\ 3727\,\AA\ line equivalent widths plotted against
redshift $z$.  Only those spectra with continuum detections near
rest-frame 3727~\AA\ better than a signal-to-noise ratio of 2 (in one
pixel) are plotted.  Uncertainty estimates are described in
Section~\ref{sec:ewsample}.  The error bars are all much smaller than
50~percent because the two-sigma limit on the continuum is a per-pixel
limit, while in fact many pixels around 3727 were used to determine
the continuum level, making the continuum measurement much more secure
than two sigma.
\label{fig:ewz}}

\plotone{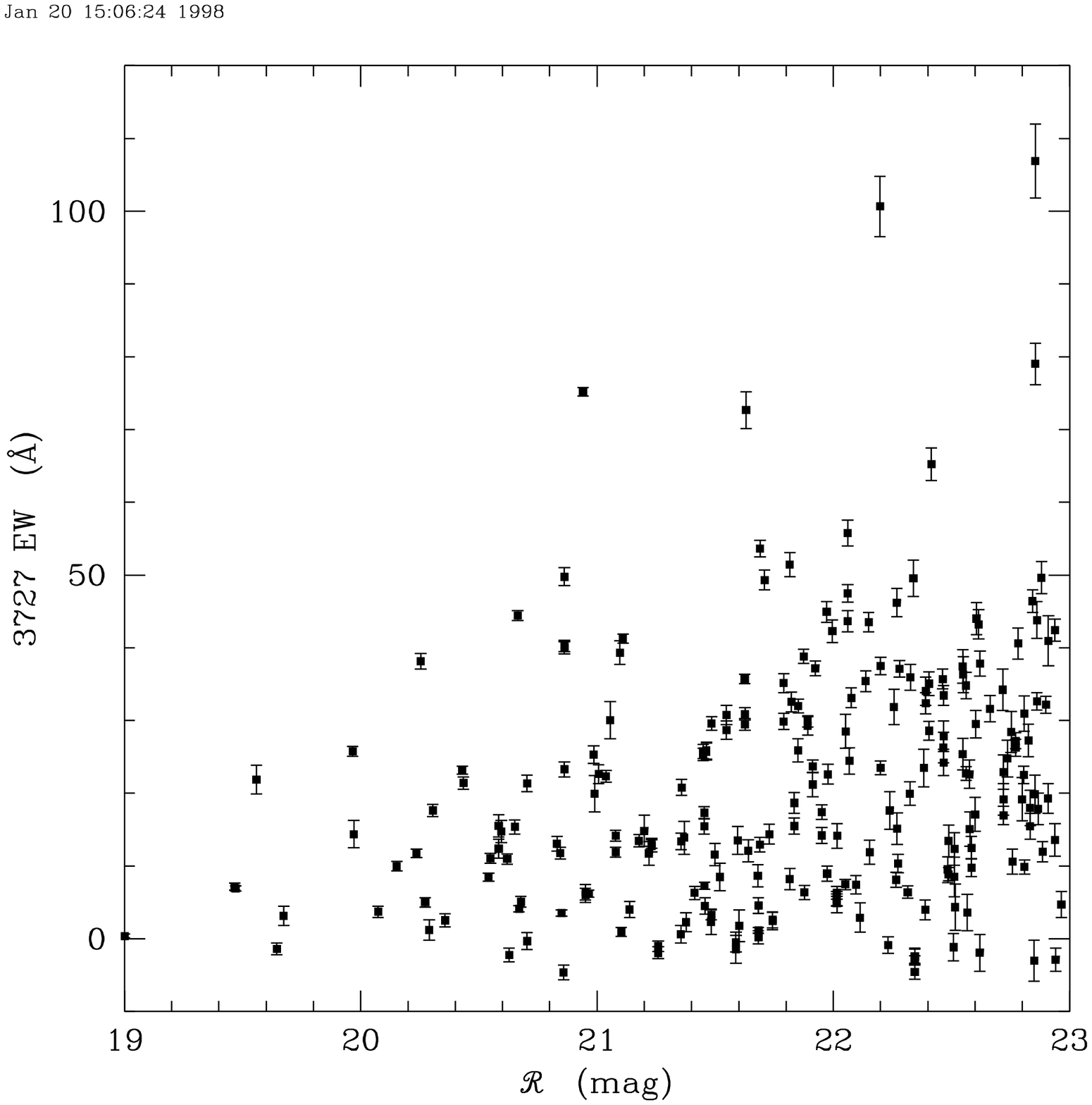}
\figcaption{
Rest-frame \OII\ 3727\,\AA\ line equivalent widths plotted against
$\cal R$-band apparent magnitude.  Only those spectra with continuum
detections near rest-frame 3727~\AA\ better than a signal-to-noise
ratio of 2 (in one pixel) are plotted.  Uncertainty estimates are
described in Section~\ref{sec:ewsample}.
\label{fig:ewR}}

\plotone{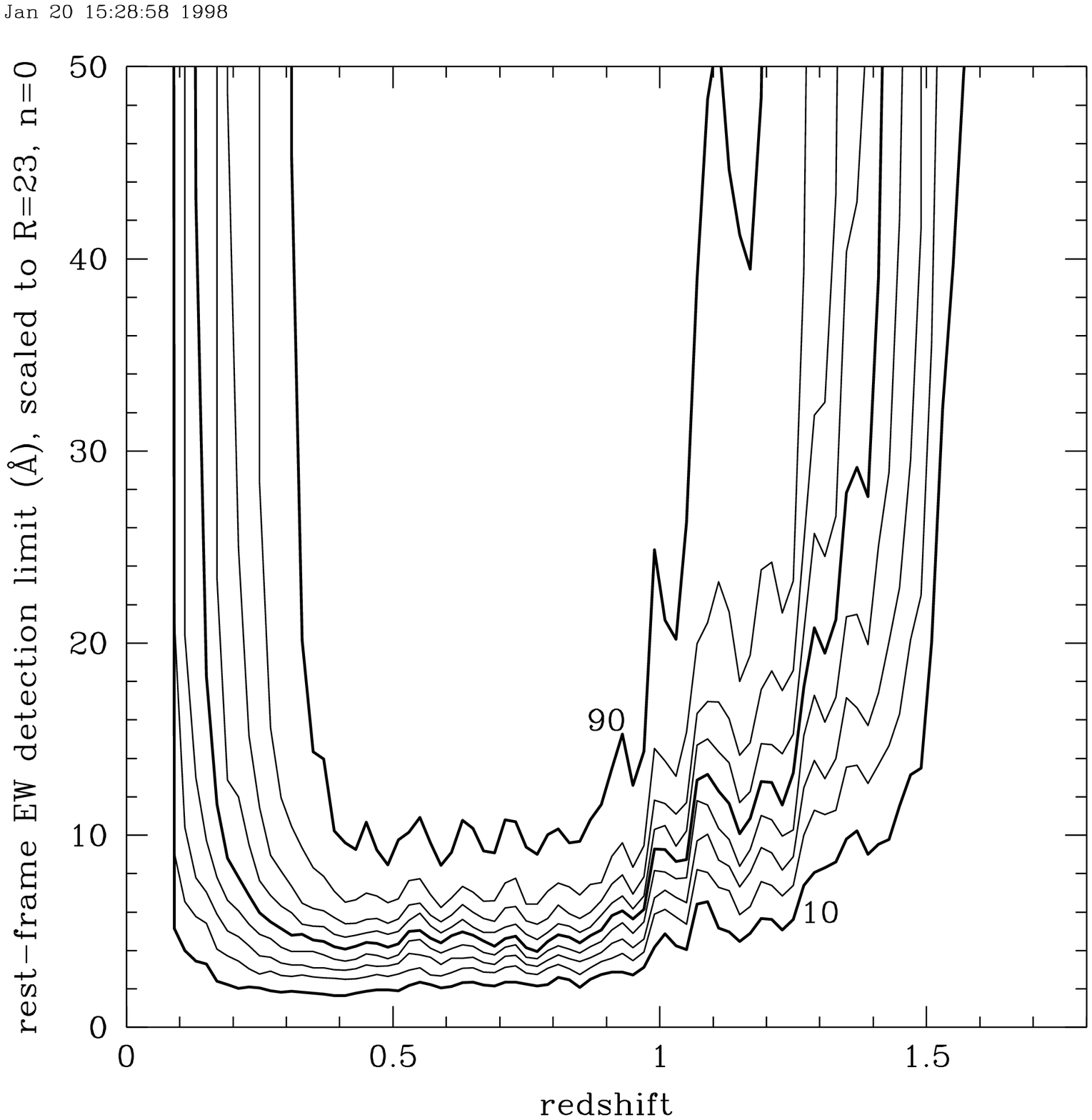}
\figcaption{
The cumulative distribution of scaled sensitivities to \OII\ 3727\,\AA
emission, in terms of rest-frame equivalent width.  The dark lines are
the 10, 50, and 90 percent contours and the thin lines are spaced by
10 percent.  The sensitivities are scaled to ${\cal R}=23$~mag and
$n=0$ (flat spectrum in $\nu f_{\nu}$) as described in the text.
\label{fig:cumulativez}}

\plotone{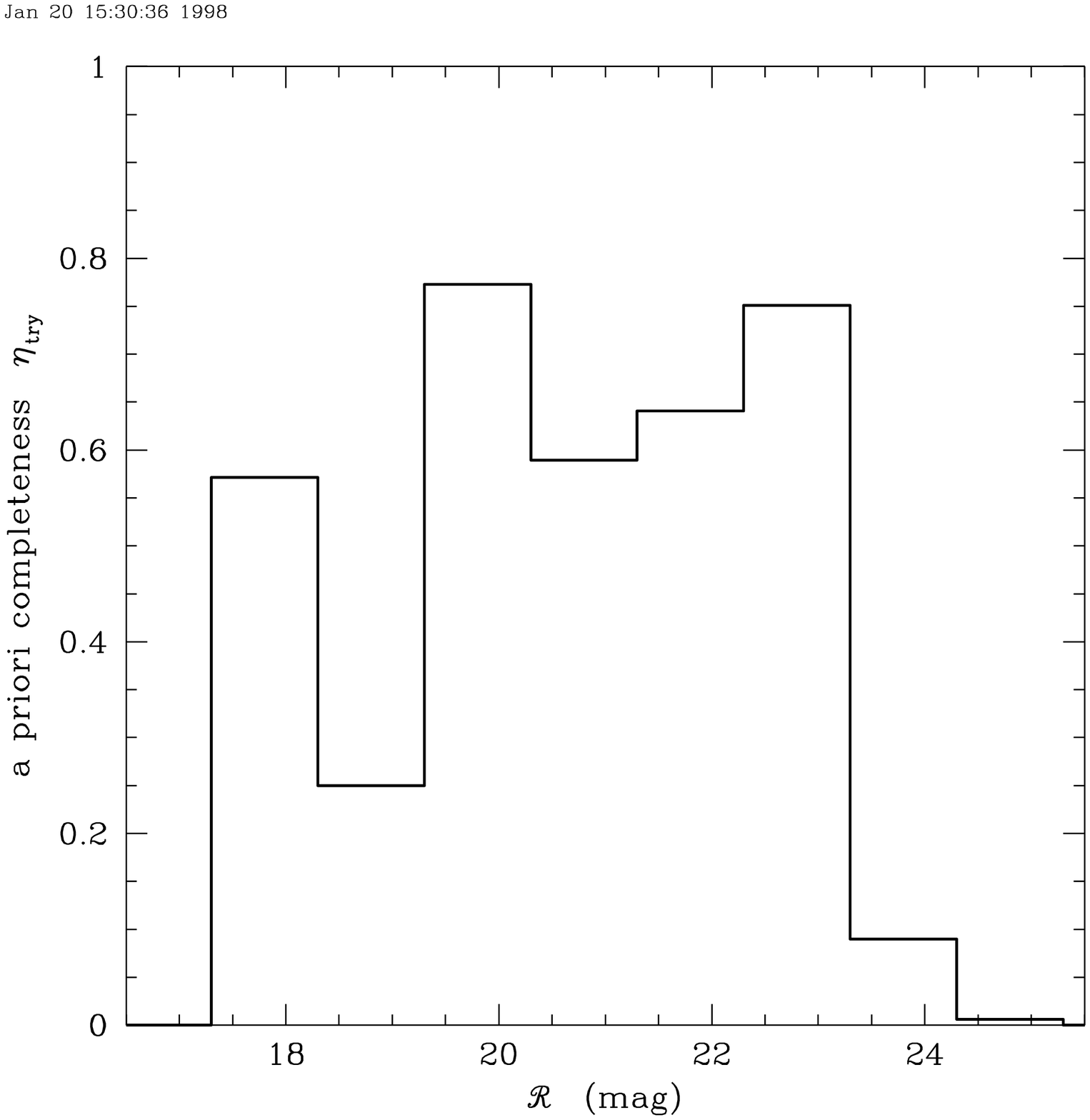}
\figcaption{
The probability, as a function of $\cal R$-band flux, that a source in
the 8-arcmin diameter HDF sample was observed spectroscopically as
part of this sample.  \Ie, this function is the fraction of sources in
the field which were observed spectroscopically.  The completeness
drops rapidly at ${\cal R}=23.3$~mag because observations fainter than
this were only performed in the central, HST-imaged, 5~$\rm arcmin^2$
of the field.
\label{fig:o2pricomp}}

\plotone{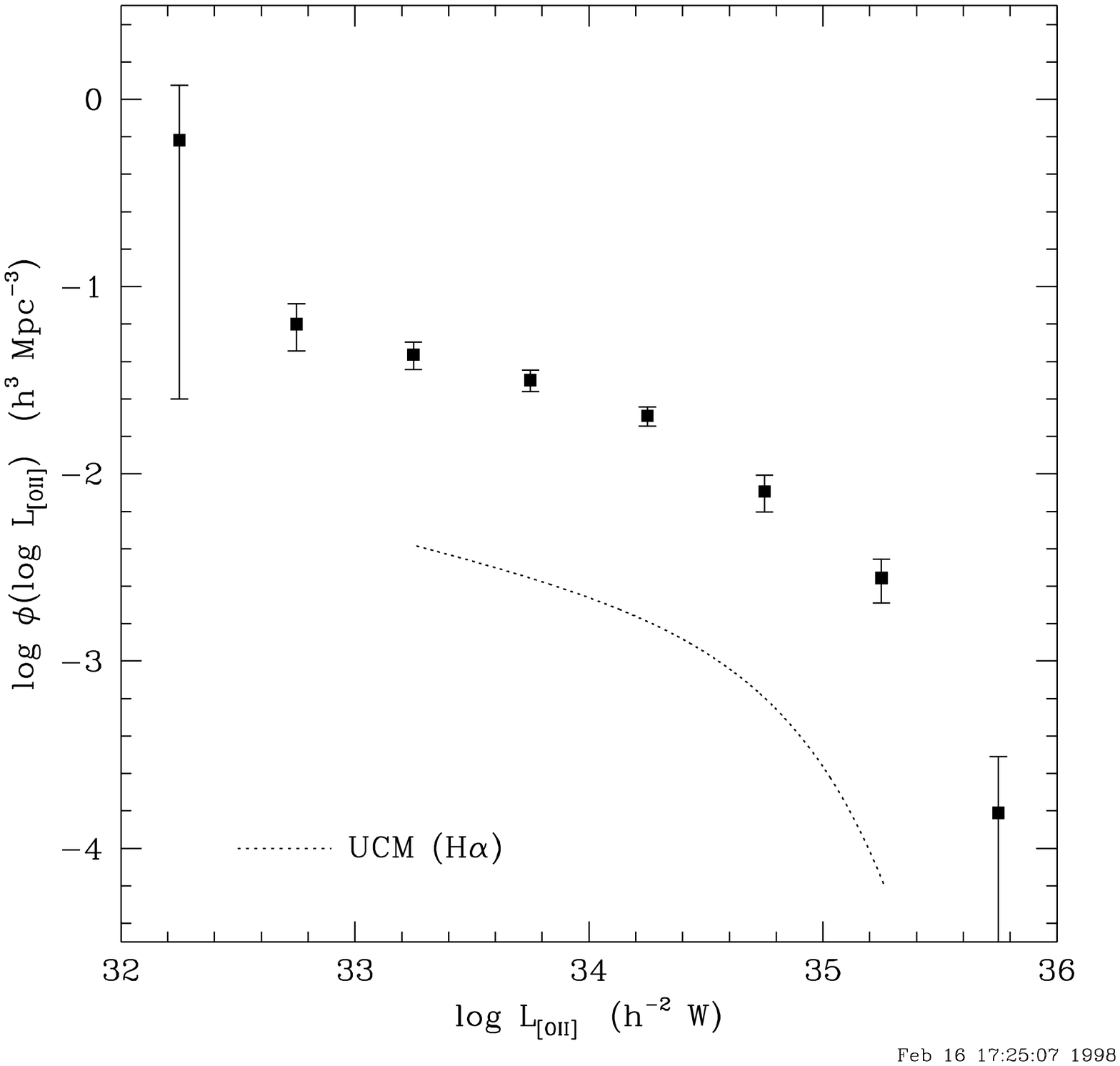}
\figcaption{
The \OII\ 3727\,\AA\ luminosity function, determined from the whole
sample used in this Chapter.  The luminosities, luminosity function
points, and uncertainties are computed as described in the text.  The
dotted line shows the Schechter-function fit to the H$\alpha$
luminosity function from the UCM survey (Gallego \etal\ 1995), in the
range in which it was determined, converted to the equivalent \OII\
luminosity function with the conversion factor $L_{\rm\OII}/L_{{\rm
H}\alpha}=0.46$, correct for the local Universe (Kennicutt 1992).
\label{fig:o2vmax0}}

\plotone{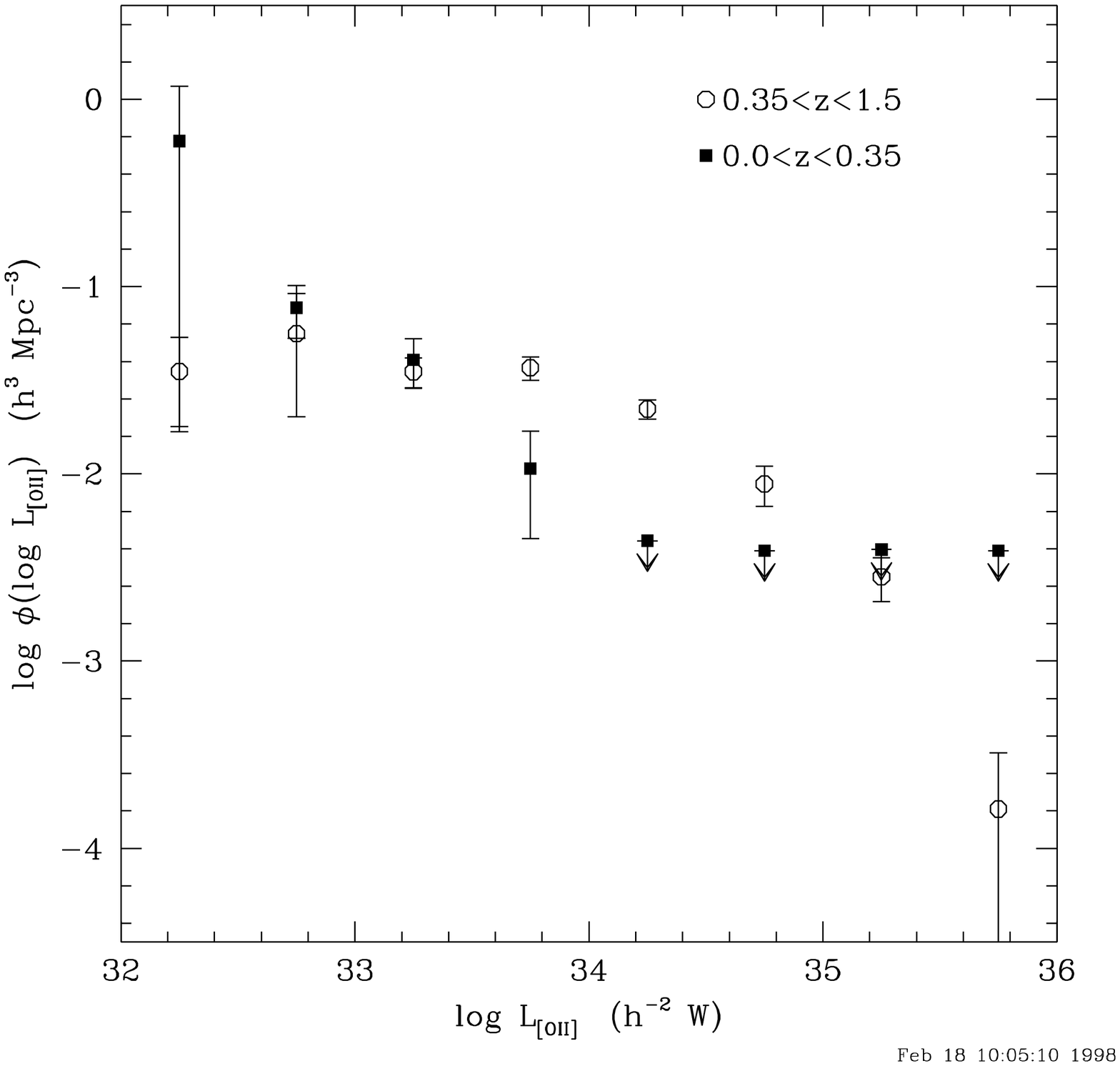}
\figcaption{
The dependence of the \OII\ 3727\,\AA\ luminosity function on
redshift, derived from two subsamples of the sample used in this
Chapter, split at redshift $z=0.35$.  The upper limit symbols show
68~percent upper limits on the low-redshift subsample, under the
conservative assumption that \OII\ line luminosity is not correlated
with broadband luminosity.
\label{fig:o2vmaxz}}

\plotone{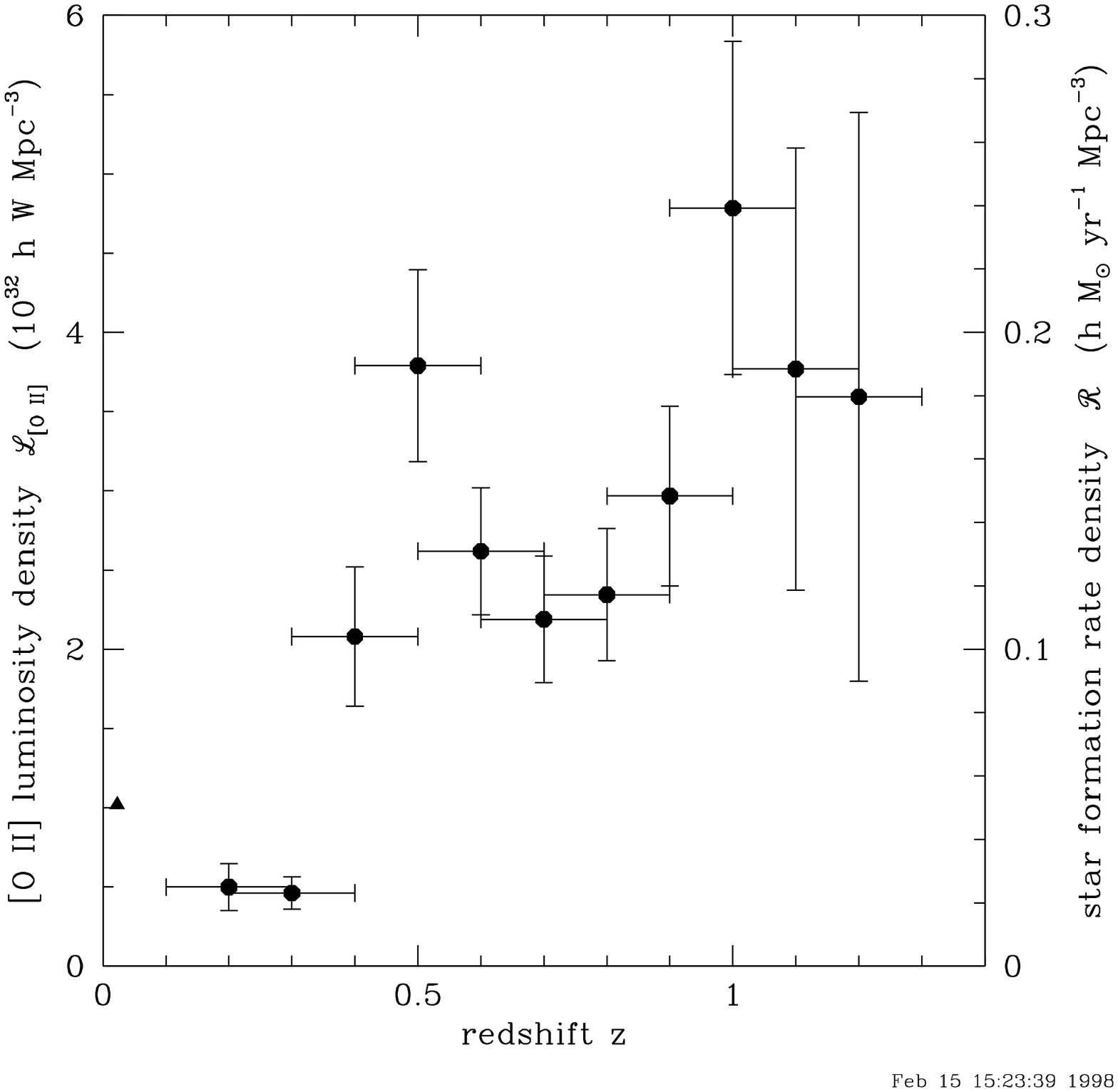}
\figcaption{
The \OII\ 3727\,\AA\ luminosity and star formation rate densities as a
function of redshift are shown with solid circles.  The luminosity
density is computed as described in the text and the star formation
rate density is computed from it using the local calibration
(Kennicutt 1992).  The horizontal error bars indicate bin widths and
the vertical error bars show poissonian uncertainties.  Note that
neighboring points are based on overlapping data; there are in effect
two independent binnings, one centered on 0.2, 0.4, \etc, and one on
0.3, 0.5, \etc\@ The solid triangle shows the local luminosity density
in the H$\alpha$ line from the UCM survey (Gallego \etal\ 1995)
converted to the equivalent \OII\ luminosity density with the
conversion factor $L_{\rm\OII}/L_{{\rm H}\alpha}=0.46$, correct for
the local Universe (Kennicutt 1992).
\label{fig:o2density}}

\end{document}